\documentclass[12pt]{l4dc2021} 

\usepackage{bm}
\usepackage{mathtools}
\usepackage{dsfont}

\makeatletter
\let\Ginclude@graphics\@org@Ginclude@graphics
\makeatother

\usepackage{textcomp}

\definecolor{coolRed}{HTML}{FF6961}
\definecolor{kindagray}{HTML}{BCBCBC}

\definecolor{bordeaux}{HTML}{D60000}
\definecolor{ripeOrange}{HTML}{ED6F38}
\definecolor{azzurro}{HTML}{0072BD}

\newtheorem{assumption}{Assumption}
\newtheorem{myRemark}{Remark}
\newtheorem{myProposition}{Proposition}

\newcommand{\hnorm}[1]{\Vert #1 \Vert_\mathcal{H}}
\newcommand{\hinner}[1]{\langle #1 \rangle_\mathcal{H}}
\DeclareMathOperator*{\argmin}{\arg\,\min}

\title[Learning-Based Model Predictive Control with Deterministic Guarantees]{KPC: Learning-Based Model Predictive Control \\ with Deterministic Guarantees}
\usepackage{times}
\author{%
 \Name{Emilio T. Maddalena$^1$} \Email{emilio.maddalena@epfl.ch}\\
 \Name{Paul Scharnhorst$^{1,2}$} \Email{paul.scharnhorst@epfl.ch}\\
 \Name{Yuning Jiang$^1$} \Email{yuning.jiang@epfl.ch}\\
 \Name{Colin N. Jones$^1$} \Email{colin.jones@epfl.ch}\\[3pt]
 \addr $^1$\'Ecole Polytechnique F\'ed\'erale de Lausanne, Lausanne, Switzerland\\
 \addr $^2$Swiss Center for Electronics and Microtechnology, Neuchâtel, Switzerland%
}

\begin{document}

\maketitle

\begin{abstract}%
We propose Kernel Predictive Control (KPC), a learning-based predictive control strategy that enjoys deterministic guarantees of safety. Noise-corrupted samples of the unknown system dynamics are used to learn several models through the formalism of non-parametric kernel regression. By treating each prediction step individually, we dispense with the need of propagating sets through highly non-linear maps, a procedure that often involves multiple conservative approximation steps. Finite-sample error bounds are then used to enforce state-feasibility by employing an efficient robust formulation. We then present a relaxation strategy that exploits on-line data to weaken the optimization problem constraints while preserving safety. Two numerical examples are provided to illustrate the applicability of the proposed control method.
\end{abstract}


\section{Introduction}

Safety is the number one requirement for any system that operates under physical constraints. For decades, this has been a major concern when control systems incorporate forms of adaptation or learning \citep{anderson2005failures,garcia2015comprehensive,hewing2020learning,wabersich2020performance}. A considerable body of literature exists establishing stability and performance guarantees in scenarios of \textit{parametric} plant-model mismatch (see \cite{lorenzen2019robust, tanaskovic2019adaptive,bujarbaruah2020exploiting} for some recent works in this direction). Depending on the final application however, assuming that the exact model structure is available might be unrealistic due to the complex physics behind the system at hand, or to the time-monetary costs associated with the modeling process.

A compelling alternative to the paradigm described above is the use of non-parametric models. These form a flexible class of surrogate functions whose number of parameters grows with the cardinality of the dataset. Relevant examples for the control community include the Nonlinear Set Memebership (NSM) \citep{milanese2004set,maddalena2020nsm} and the Kinky Inference (KI) \citep{calliess2020lazily,manzano2020robust} techniques. Due to the ease of incorporating prior expert knowledge and the inherent uncertainty quantification associated with them, Gaussian processes (GPs) have recently become a popular modeling tool for dynamical systems \cite{capone2020localized,matschek2020learningA,arcari2020meta,umlauft2020learning,shukla2020convergence,yingzhao2020gaussian}. Such function approximators are typically paired with appropriate Model Predictive Control (MPC) schemes that not only take into account the latent function estimate, but also the model variance to act with care in highly uncertain regions of the space (see \cite{koller2018learning,hewing2019cautious} for two examples, and \cite{beckers2019stable} for an exception to this trend).

As opposed to Gaussian processes, kernel ridge regression (KRR) and support vector regression (SVR) are deterministic non-parametric tools. These models have the same form of a GP predictive mean: a weighted sum of kernel basis functions. Moreover, with an appropriate choice of regularization constant, a KRR model matches exactly a GP posterior \citep{kanagawa2018gaussian}. Connections between the stochastic and the deterministic frameworks are profound and have been long known \citep{kimeldorf1970correspondence}. Uncertainty can be quantified in the KRR and SVR cases by considering all maps belonging to their underlying reproducing kernel Hilbert space (RKHS) of functions. The RKHSs associated to various kernels, including the widely used squared-exponential, are dense in the space of continuous functions with compact domains \citep{micchelli2006universal}.

\textbf{Our contribution:} We propose in this paper a predictive control strategy based on non-parametric kernel regression that incorporates deterministic guarantees of safety. Samples from the unknown ground-truth dynamics are used to construct one-step and multi-step ahead models, with an appropriate state-dependent uncertainty quantification obtained from recently derived error-bounds \citep{maddalena2020deterministic}. The available dataset can be contaminated by noise, which is only assumed to be bounded, but otherwise drawn from any distribution. An efficient robust optimization formulation is derived to enforce state-constraint satisfaction. We then present a relaxation strategy that exploits on-line information to alleviate the problem constraints while preserving safety. Two numerical examples are provided and we discuss scalability issues to large datasets.

\section{Problem Definition}

We consider a discrete-time nonlinear system of the form
\begin{equation}
    x_{t+1} = f(x_t, u_t)
    \label{eq.groundTruth}
\end{equation}
with time $t \in \mathbb{N}$, states $x_t \in \mathbb{R}^{n_x}$, inputs $u_t \in \mathbb{R}^{n_u}$, and \textit{unknown}\footnote{If a partial model for the latent function is available, the learning task is simply performed on the error dynamics.} transition map $f: \mathbb{R}^{n_x} \times \mathbb{R}^{n_u} \rightarrow \mathbb{R}^{n_x}$. Hard constraints $x_t \in \mathbb{X} = \{x \, | \, g_i(x) \leq 0, i=1,\dots,n_{\mathbb{X}}\}$ and $u_t \in \mathbb{U} = \{u \, | \, s_i(u) \leq 0, i=1,\dots,n_{\mathbb{U}}\}$ are imposed for all $t \in \mathbb{N}$, where $\mathbb{X}$ and $\mathbb{U}$ are polyhedra. More general compact sets could also be considered herein, but the geometric assumptions posed on $\mathbb{X}$ and $\mathbb{U}$ will later allow for an efficient robust reformulation as detailed in Section~\ref{sec.KPC}. Given a \textit{safe subset} of the state space
\begin{equation}
    \mathbb{X}_{\text{safe}} \subseteq \mathbb{X}
\end{equation}
our goal is to drive the dynamical system \eqref{eq.groundTruth} from a specified initial condition $x_0$ to the set $\mathbb{X}_{\text{safe}}$ while satisfying all constraints. Similarly to \cite{koller2018learning}, we also assume that a local policy $\pi_{\text{safe}}:\mathbb{X}_{\text{safe}} \rightarrow \mathbb{U}$ is available, making $\mathbb{X}_{\text{safe}}$ forward invariant, i.e., $\forall x_t \in \mathbb{X}_{\text{safe}}: \, f(x_t,\pi_{\text{safe}}(x_t)) \in \mathbb{X}_{\text{safe}}, \pi_{\text{safe}}(x_t) \in \mathbb{U}$. A frequent instance of this problem is the regulation to a specific fixed point, in which $\mathbb{X}_{\text{safe}} = \{x_{\text{eq}}\}$ and $\pi(x_{eq}) = u_{eq}$ is the equilibrium control constant. In order to accomplish our task, we make use of noise-corrupted measurements of our unknown ground-truth and the formalism of non-parametric kernel learning are described next.

\section{Non-Parametric Kernel Learning}
In this section only, we consider $n_x = 1$ simply to avoid using a cumbersome notation; for $n_x > 1$, each output component of $f$ has to be considered separately. Moreover, the shorthand notation $f(x,u) = f(z)$ is used. Suppose the map $f$ is unknown, but a collection of $D$ measurement pairs is available to reconstruct it
\begin{equation}
    \mathbb{D} = \{(z_d, y_d) \}_{d=1}^{D} \quad \text{where} \quad y_d = f(z_d) + \delta_d
\end{equation}
We make the following two assumptions on our dataset and on the observational model.

\begin{assumption}
\label{as.pairwise}
The data locations $z_1, \dots, z_D$ are pairwise distinct.
\end{assumption}

\begin{assumption}
\label{as.noise}
The noise affecting all data-points $\delta = \left( \delta_1 \, \dots \, \delta_D \right)$ is bounded in module by a known quantity $\bar{\delta } \geq |\delta|$, $\bar{\delta} \in \mathbb{R}^D_{\geq0}$.
\end{assumption}

The approach of kernel machines is employed to learn the unknown dynamics from the available dataset. Next, we recall the basics of such theory, see \cite{schlkopf2018learning} for a more complete coverage of the topic. A kernel is any real-valued symmetric positive-semidefinite function $k: \mathcal{Z} \times \mathcal{Z} \rightarrow \mathbb{R}$. Each kernel defines a reproducing kernel Hilbert space $\mathcal{H} \subset \mathbb{R}^{\mathcal{Z}}$, where $\forall z \in \mathcal{Z}$ we have that $k(z,\cdot) \in \mathcal{H}$. Computing the inner-product between a map $h \in \mathcal{H}$ and a partially evaluated kernel $k(z,\cdot)$ is equivalent to assessing the value of $h$ at $z$, i.e., $\langle h,k(z,\cdot) \rangle_\mathcal{H} = h(z)$, which is known as the reproducing property. Members $f$ of $\mathcal{H}$ can be seen as linear combinations of partially evaluated kernel functions since $\mathcal{H}$ is the closure of $\text{span}(k(z,\cdot)), \forall z \in \mathcal{Z}$ with respect to the induced metric. The norm in the $\mathcal{H}$ space is defined as $\hnorm{h} = \sqrt{\hinner{h,h}}$.

For convenience, we define $Z$ as the collection of all dataset inputs $z_d$, and $y$ as the collection of all targets $y_d$. Also, let $K \in \mathbb{R}^{D \times D}$ be the constant matrix of kernel evaluations at $Z$, i.e., $k(z_i,z_j)$ at its i-th row and j-th column, and let $K_{Zz} : \mathcal{Z} \rightarrow \mathbb{R}^D$ denote the column vector \textit{function} $z \mapsto \left(k(z_1,z), \dots, k(z_D,z)\right)^\top$ and $K_{zZ}$ simply represents its transpose. Finally, the so-called power function is a non-negative map $P:\mathcal{Z}\rightarrow \mathbb{R}_{\geq 0}$ defined as
\begin{equation}
    P(z) = \sqrt{k(z,z) - K_{zZ}K^{-1}K_{Zz}}
\end{equation}
and evaluates to zero for all $z_d$ in the dataset \citep[Sec. 11.1]{wendland2004}. Note the similarity between the power function $P(z)$ and the posterior variance of a Gaussian process.

The estimate $\hat f$ is built by minimizing a combination of the mean-squared error and a regularization term to penalize complexity, i.e., the kernel ridge regression (KRR) cost
\begin{equation}
    \hat f = \argmin_{h \in \mathcal{H}} \left \{ \frac{1}{D} \sum_{(z,y) \in \, \mathbb{D}}  \, (h(z) - y)^2 + \lambda \, \hnorm{h}^2 \right \}.
\end{equation}
According to the well-known representer theorem \citep{schlkopf2018learning}, out of all possible maps $h \in \mathcal{H}$, a minimizer exists and is given by a weighted sum of kernels centered at the input locations $Z$. The problem above is therefore equivalent to a finite-dimensional quadratic program whose closed-form solution, our nominal model, is given by
\begin{equation}
    \label{eq.KRR}
    \hat f =  K_{zZ} \, (K+D\lambda I)^{-1} y
\end{equation}
%in which the vector of weights is entirely determined by the data, the kernel matrix and the regularizartion constant: $\alpha = (K + D \lambda I)^{-1} y$.

\begin{myRemark}
\normalfont Note that the map described by \eqref{eq.KRR} has the same form as a Gaussian process posterior distribution conditioned on the data, that is, its predictive mean. Indeed, if $\sigma$ is the noise variance in the GP scenario and $\lambda$ is selected as $\sigma^2 / D$, the two models are exactly the same. The reader is referred to \cite[\textsection~6.1]{williams2006gaussian} for a discussion on the existing connections.
\end{myRemark}

\begin{assumption}
\label{as.kernel}
The chosen kernel $k(\cdot, \cdot)$ is a \textit{strictly} positive-definite function.
\end{assumption}

\begin{assumption}
\label{as.ground}
The unknown dynamics $f$ are contained in the RKHS of the chosen kernel $k(\cdot,\cdot)$, and an upper bound for its norm is available $\Gamma \geq \hnorm{f}$.
\end{assumption}

The two conditions above are central to the development of the control strategy safety guarantees. Assumption~\ref{as.kernel} can be satisfied by selecting an appropriate kernel function such as the squared-exponential or the inverse multiquadrics. Assumption~\ref{as.ground} encapsulates our knowledge about the complexity of the unknown ground-truth: intuitively, the more kernel basis functions are needed to describe it, the larger the associated norm. The same piece of information is required in the works \cite{koller2018learning,hashimoto2020learning} as well as in various other recent papers. In \cite{maddalena2020deterministic}, an example is provided on how $\Gamma$ could be estimated from noiseless samples of the latent function, and how this estimation process is affected by the presence of bounded noise. As shown in the latter work, finite-sample deterministic error bounds exist for KRR models.

\begin{theorem}[{\normalfont \cite{maddalena2020deterministic}}]
Let $K$ be the kernel matrix, $D$ be the number of data-points, $\bar{\delta} \in \mathbb{R}^D_{\geq0}$ the noise bound, and $\lambda > 0$ be the regularization constant. Under Assumptions~\ref{as.pairwise} to \ref{as.ground}, the KRR model $\hat f$ admits the following prediction error bound for any $z \in \mathcal{Z}$
\begin{equation}
\left|\hat{f}(z)-f(z)\right| \leq \, P(z) \sqrt{ \Gamma^2 + \Delta - y^\top K^{-1} y} \, + \bar{\delta}^\top \, \vert K^{-1}K_{Zz} \vert + \left| \, y^\top \left(K + \frac{1}{D \lambda} \, KK \right)^{-1} K_{Zz}\right|
\label{eq.boundKRR}
\end{equation}
where $f$ is the unknown ground-truth and $\Delta = \max \{- \delta^\top K^{-1} \delta$ $+2 y ^\top K^{-1} \delta \}$ subject to $|\delta| \leq \bar{\delta}$.
\label{th.mainOne}
\end{theorem}

Notice that the bound above can be easily evaluated, and the only term that is not given in closed-form is the constant $\Delta$--- which requires solving a box-constrained quadratic program over $D$ variables. This quantity is independent of the query point $z$ and compensates for a possible underestimation of the model complexity caused by the noise \citep[Lemma~1]{maddalena2020deterministic}. If one wishes not to solve such an optimization problem, then the entire square-root term could be replaced by $\Gamma$ at the expense of increasing the bound conservativeness. The term $\bar{\delta}^\top \, \vert K^{-1}K_{Zz} \vert$ accounts for the potentially adversarial nature of $\delta$ and, finally, the last term penalizes the use of high regularization constants $\lambda$. For a stable numerical evaluation of \eqref{eq.boundKRR}, a small diagonal \textit{jitter} has to be added to the Gram matrix as customary in the field of Gaussian processes \citep{bauer2016understanding}.

\begin{myRemark}
\normalfont
When compared to the GP bounds presented in \cite{srinivas2012information}, the result given in Theorem~\ref{th.mainOne} does not involve information-theoretic measures such as the maximal information gain. The need of estimating such constant hampers the applicability of the former bounds in practical scenarios (see the discussion in \cite{lederer2019uniform}). When compared to the results in \cite[Lemma~2]{hashimoto2020learning}, the inequality \eqref{eq.boundKRR} tends to give rise to tighter bounds as shown in \cite{maddalena2020deterministic}; nevertheless, the latter are more unstable at the extremes of the input space. In order to avoid this effect, data have to ideally fill the ground-truth domain while still being well-separated. In the approximation theory community, this interplay between precision and stability is known as the uncertainty principle \citep{wendland2004}.
\end{myRemark}

\section{Kernel Predictive Control}
\label{sec.KPC}

One possible approach to tackling our problem is to build a single-step surrogate model for the latent function and employ uncertainty propagation techniques to perform multi-step ahead predictions. Propagating sets through general non-linear maps is challenging and usually involves several overbouding steps \citep{koller2018learning}. Therefore, we opt for learning various condensed models, one for each of the $N$ prediction steps. Let $F_1: \mathbb{X} \times \mathbb{U} \rightarrow \mathbb{X}$ be the one-step ahead predictor in which each dimension in learned separately by KRR models $\hat f_1, \dots, \hat f_{n_x}$
\begin{equation}
    (x_0,u_0) \mapsto F_1(x_0,u_0) = \left( \hat{f}_1(x_0,u_0), \dots, \hat{f}_{n_x}(x_0,u_0)\right)
\end{equation}
Define $F_2$ as $(x_0,u_0,u_1) \mapsto F_2(x_0,u_0,u_1) = \left( \hat{f}_1(x_0,u_0,u_1), \dots, \hat{f}_{n_x}(x_0,u_0,u_1)\right)$, the two-step ahead model, and $F_3,\dots,F_N$ analogously. Robust \textit{confidence sets} are then built around our nominal predictions. Let $\mathcal{X}_1: \mathbb{X} \times \mathbb{U}: \rightarrow 2^{\mathbb{X}}$ be a set-valued function defined as the hyper-rectangle
\begin{equation}
    (x_0,u_0) \mapsto \mathcal{X}_1(x_0,u_0) = \left(\hat{f}_{1}(x_0,u_0) \pm \beta_1(x_0,u_0), \, \dots, \, \hat{f}_{n_x}(x_0,u_0) \pm \beta_{n_x}(x_0,u_0) \right)
    \label{eq.setValued}
\end{equation}
where $\beta(x,u)$ denotes the right-hand side of the inequality \eqref{eq.boundKRR}, and $a \pm b$ refers to the set $\{c \, | \, a-b \leq c \leq a+b\}$. Similarly, define also the maps $\mathcal{X}_2,\dots\mathcal{X}_N$, which share the same domain respectively with $F_2,\dots,F_N$. As a direct consequence of Theorem~\ref{th.mainOne}, we have that
\begin{equation}
    \begin{aligned}
    \mathcal{X}_1(x_0,u_0) &\ni f(x_0,u_0) \\
    \mathcal{X}_2(x_0,u_0,u_1) &\ni f(f(x_0,u_0),u_1) \\
    & \dots \\
    \mathcal{X}_N(x_0,u_0,u_1,\dots,u_{N-1}) &\ni f(\dots f(f(x_0,u_0),u_1),\dots,u_{N-1}).
    \end{aligned}
    \label{eq.setValueds}
\end{equation}

\begin{figure}[b!]%[htbp]
    \floatconts
    {fig:example}% label
    {\vspace{-18pt} \caption{A schematic representation of the outcome of each model. Nominal predictions $F_t$ (\textbf{|}), confidence sets $\mathcal{X}_t$ (\textbf{- -}), and the unknown ground-truth $f$ (\textcolor{kindagray}{\textbf{|}}). All functions also depend on the chosen control sequence, which is omitted for clarity.}}% 
    {\includegraphics[width=0.78\textwidth]{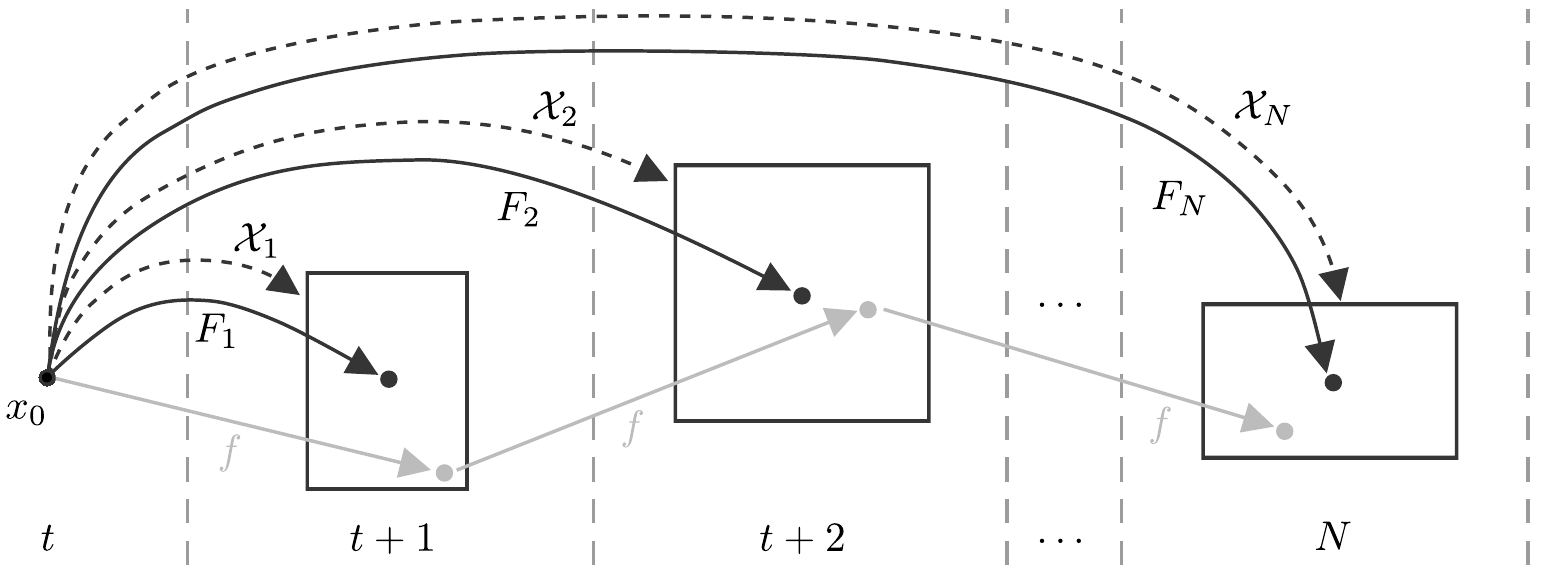}}
\end{figure}

\begin{myRemark}
\normalfont
Training the associated kernel models requires various $N$-step experiments to be performed rather than simply having one-step ones. For instance, the model $F_N$ requires multiple tuples $(x_0,u_0,\dots,u_{N-1})$ as features and (possibly noisy) measurements of the resulting states $x_N$ as targets. We highlight that long sequences of linked states, i.e., long experiments, are preferred over various short ones even in classical parametric system identification \citep{lennart1999system}.
\end{myRemark}
%and $\hat{f}_1, \beta_1, \dots, \hat{f}_{x_n}, \beta_{n_x}$ are non-parametric real-valued kernel regressors and their bounds built for each state dimension.

Let $x_0$ be a \textit{given} initial condition for the true dynamical system \eqref{eq.groundTruth}. Our Kernel Predictive Control formulation is expressed as the finite-horizon optimal control problem
\begin{subequations}
\begin{align}
    \text{KPC}: \; \min_{X,U} & \quad \sum_{t=0}^{N-1} \ell (x_t,u_t) + \ell_f (x_N) \\
    \text{s.t.} & \quad x_{t} = F_{t}(x_0,u_0,\dots,u_{t-1}), \, \forall t \label{eq.KPCnominal} \\
                & \quad \mathcal{X}_t(x_0,u_0,\dots,u_{t-1}) \subseteq \mathbb{X}, \; \forall t \hspace{2cm} \label{eq.setContrs1} \\
                & \quad \mathcal{X}_N(x_0,u_0,\dots,u_{N-1}) \subseteq \mathbb{X}_{\text{safe}} \label{eq.setContrs2} \\
                & \quad u_t \in \mathbb{U}, \, \forall t 
\end{align}
\label{eq.optControlProb}%
\end{subequations}
where $X = (x_1,\dots,x_N)$, $U = (u_0,\dots,u_{N-1})$ are the decision variables, and $\ell(x,u)$ and $\ell_f(x)$ are appropriately designed stage and final costs. In a receding-horizon implementation, KPC is solved recursively and only the first optimal control inputs are applied to the system.

\begin{myProposition}
Let the \emph{KPC} problem \eqref{eq.optControlProb} be feasible and $(X^\star,U^\star)$ be any of its feasible solutions. The sequence of inputs $U^\star = (u_0^\star, \dots, u_{N-1}^\star)$ drives the true system \eqref{eq.groundTruth} from $x_0$ to the safe set $\mathbb{X}_{\text{safe}}$ while satisfying the constraints at all times, i.e., $f(x_0,u_0^\star), \, f(f(x_0,u_0^\star),u_1^\star), \dots \, \in \mathbb{X}$, and $f(\dots f(f(x_0,u_0^\star),u_1^\star),\dots,u_{N-1}^\star) \in \mathbb{X}_{\text{safe}}$.
\label{prop1}
\end{myProposition}

\noindent \textbf{Proof:} Follows from the definition of the sets \eqref{eq.setValued}, the validity of the deterministic bounds \eqref{eq.boundKRR}, and the imposed constraints \eqref{eq.setContrs1} and \eqref{eq.setContrs2}.
\bigbreak

The KPC constraints \eqref{eq.setContrs1} and \eqref{eq.setContrs2} are set inclusions that need to be reformulated to allow for numerical computations. In what follows, we employ a coordinate transformation and exploit the closed-form solution of the obtained hyper-cube support function. Let $H_i$ and $h_i$ be the parameters of the $i$th half-space of $\mathbb{X}$ and consider the condition $\mathcal{X}_t \subseteq \mathbb{X}$, where the arguments of $\mathcal{X}_t$ are omitted to ease notation. This constraint is satisfied if for each one of the half-spaces $g_i(x) \leq 0$ that describe the polyhedron $\mathbb{X}$ it holds that
\begin{align}
     & \forall x \in \mathcal{X}_t: \, g_i(x) \leq 0 \ \\[3pt]
     \Leftrightarrow \, & \max \{ g_i(x) \, | \, x \in \mathcal{X}_t \} \, \leq 0 \\[3pt]
     \Leftrightarrow \, & \max \{ H_i^\top \, x - h_i \, | \, \vert x - F_t \vert \leq \mathcal{B}_t \} \, \leq 0 \label{eq.constr} \\[3pt]
     %
     %\Leftrightarrow \, & \max \{ H_i \, \Omega_t (x+F_t) - h_i \, | \, \Vert x \Vert_\infty \leq 1 \} \, \leq 0 \\[3pt]
     %
     %\Leftrightarrow \, & \, \Vert H_i \, \Omega_t \Vert_1 + H_i \, \Omega_t F_t - h_i \, \leq 0 \label{eq.withOneNorm}
     \Leftrightarrow \, & \max \{ H_i^\top \, (B_t \, x+F_t) - h_i \, | \, \Vert x \Vert_\infty \leq 1 \} \, \leq 0 \label{eq.withInfNorm} \\[3pt]
     \Leftrightarrow \, & \, \Vert H_i^\top B_t \Vert_1 + H_i^\top F_t - h_i \, \leq 0 \label{eq.withOneNorm} 
     %\\[3pt]
     %
     %\Leftrightarrow \, & \, \exists \lambda_i: \, \vert H_i \, \Omega_t \vert - \lambda_i \leq 0, \ \, \textbf{1}^\top \lambda_i + H_i \, F_t - h_i \, \leq 0
\end{align}
where $F_t$ and $\mathcal{B}_t = (\beta_1,\dots,\beta_{n_x})^\top \in \mathbb{R}^{n_x}$ are the parameters of the confidence set $\mathcal{X}_t$ (see \eqref{eq.setValued}), and $B_t = \text{diag}\left( \beta_1,\dots,\beta_{n_x} \right) \in \mathbb{R}^{n_x \times n_x}$. Lastly, the one-norm can be eliminated from \eqref{eq.withOneNorm} by introducing new auxiliary variables and inequality constraints, a standard linear programming procedure.

\begin{myRemark}
\normalfont
The maximization in \eqref{eq.constr} could also be directly converted into its dual form without the reformulation \eqref{eq.withInfNorm}. Although this would not introduce any conservatism, additional decision variables would be created along with \textit{nonlinear equality constraints}, thus significantly increasing the KPC formulation complexity. The approach adopted above is both economic and exact.
\end{myRemark}

After converting the set constraints \eqref{eq.setContrs1} and \eqref{eq.setContrs2} into the form \eqref{eq.withOneNorm}, one obtains
\begin{subequations}
\begin{align}
     \; \min_{X,U} & \quad \sum_{t=0}^{N-1} \ell (x_t,u_t) + \ell_f (x_N) \\[3pt]
    \text{s.t.} & \quad x_{t} = F_{t}(x_0,u_0,\dots,u_{t-1}), \, \forall t \\[3pt]
                %& \quad \Vert H_i \, \Omega_t(x_0,U_{[0, \, t-1]}) \Vert_1 + H_i \, \Omega_t(x_0,U_{[0, \, t-1]}) F_t(x_0,U_{[0, \, t-1]}) - h_i \, \leq 0,  \, \forall i \label{in.state} \\
                %& \quad H_i \, \Omega_t(x_0,u_0,\dots,u_{t-1}) F_t(x_0,u_0,\dots,u_{t-1}) + \textbf{1}^\top\lambda_{t,i} - h_i \, \leq 0,  \, \forall i \label{in.runing} \\
                %& \quad \vert H_i \, \Omega_t(x_0,u_0,\dots,u_{t-1}) \vert - \lambda_{t,i} \, \leq 0, \, \forall i \\[3pt]
                %
                %& \quad Q_i \, \Omega_N(x_0,u_0,\dots,u_{N-1}) F_N(x_0,u_0,\dots,u_{N-1}) + \textbf{1}^\top\lambda_{N,i} - q_i \, \leq 0,  \, \forall i \label{in.terminal} \\
                %& \quad \vert Q_i \, \Omega_N(x_0,u_0,\dots,u_{N-1}) \vert - \lambda_{N,i} \, \leq 0, \, \forall i \label{in.last} \\[3pt]
                & \quad \Vert H_i^\top B_t(x_0,u_0,\dots,u_{t-1}) \Vert_1 \, + H_i^\top F_t(x_0,u_0,\dots,u_{t-1}) - h_i \, \leq 0,  \, \forall i, \forall t \label{in.runing} \\[3pt]
                & \quad \Vert Q_i^\top B_N(x_0,u_0,\dots,u_{N-1}) \Vert_1 \, + Q_i^\top F_N(x_0,u_0,\dots,u_{N-1}) - q_i \, \leq 0,  \, \forall i \label{in.last} \\[3pt]
                & \quad S \, u_t \leq s, \, \forall t
\end{align}
\label{eq.KPC2}
\end{subequations}
in which $H_i, Q_i \in \mathbb{R}^{n_x}$, $h_i, q_i \in \mathbb{R}$ for all $i$, $S \in \mathbb{R}^{n_\mathbb{U} \times n_u}$ and $s \in \mathbb{R}^{n_\mathbb{U}}$. Finally, if a solution to \eqref{eq.KPC2} is found, the true system can be steered to $\mathbb{X}_{\text{safe}}$ in open-loop; however, this feasibility does not guarantee that future iterations of the same problem will also be feasible.

\subsection{A Safe Relaxation Strategy}

Next we propose a safe relaxation strategy (SRS) that can be used to weaken the optimization problem constraints \eqref{in.runing}-\eqref{in.last} whenever data from previous KPC iterations are available. 

\begin{myProposition}
Assume that the $N-1$ previous consecutive \emph{KPC} iterations were feasible. Denote by $(u_{-N+1}, \dots, u_{-1})$ and $(x_{-N+1}, \dots, x_{-1})$ the closed-loop sequences of past controls and states, and by $x_0$ the current state. Let \emph{KPC}$^+$ be the \emph{KPC} optimization problem \eqref{eq.optControlProb} with set constraints $\mathcal{X}_1(x_0,u_0) \subseteq \mathbb{X}$, $\mathcal{X}_2(x_0,u_0,u_1) \subseteq \mathbb{X}$, $\dots$, $\mathcal{X}_{N-1}(x_0,u_0,\dots,u_{N-2}) \subseteq \mathbb{X}$ relaxed to
\begin{equation}
\label{eq.SRS}
\begin{aligned}
\mathcal{X}_1(x_0,u_0) &\cap \left( \, \bigcap_{i = 2}^{N} \mathcal{X}_{i} \, (x_{-i+1},u_{-i+1},\dots,u_{-1},u_0) \right) \subseteq \mathbb{X}, \\
\mathcal{X}_2(x_0,u_0,u_{1}) &\cap \left( \, \bigcap_{i=3}^N \mathcal{X}_{i} \, (x_{-i+2},u_{-i+2},\dots,u_{-1},u_0,u_1) \right) \subseteq \mathbb{X}, \ \dots
\end{aligned}
\end{equation}
Let \emph{KPC}$^+$ be feasible and $(X^\star,U^\star)$ be any of its feasible solutions. Then, $U^\star = (u_0^\star, \dots, u_{N-1}^\star)$ drives the true system \eqref{eq.groundTruth} from $x_0$ to the safe set $\mathbb{X}_{\text{safe}}$ while satisfying the constraints at all times, i.e., $f(x_0,u_0^\star), \, f(f(x_0,u_0^\star),u_1^\star), \dots \, \in \mathbb{X}$, and $f(\dots f(f(x_0,u_0^\star),u_1^\star),\dots,u_{N-1}^\star) \in \mathbb{X}_{\text{safe}}$.
\label{prop.SRS}
\end{myProposition}

\noindent \textbf{Proof:} %From construction, we know that $\mathcal{X}_t (x_0,u_0,\dots,u_{t-1}) \ni f()$ for any $t$. If the previous KPC iteration was feasible, then also $\mathcal{X}_{t+1} (x_{-1},u_{-1},u_0,\dots,u_{t-1}) \ni x_t$, where $x_{-1}$ and $u_{-1}$ are past closed-loop data. Considering a total of $N$ previous consecutive feasible KPC iterations yields a total of $N$ set conditions for $x_1$, $N-1$ set conditions for $x_2$, $\dots$, $2$ set conditions for $x_{N-1}$, and $1$ set condition for $x_N$. The states are therefore contained in the intersection of all their associated sets and, thus, enforcing the relaxed constraints suffices to ensure constraint satisfaction.
From construction, we know that $\mathcal{X}_t (x_0,u_0^\star,\dots,u_{t-1}^\star) \ni f(\dots f(f(x_0,u_0^\star),\dots),u_{t-1}^\star)$ for any $t$. If the previous KPC iteration was feasible, then also $\mathcal{X}_{t+1} (x_{-1},u_{-1},u_0^\star,\dots,u_{t-1}^\star)$ contains the same point, where $x_{-1}$ and $u_{-1}$ are past closed-loop data. Considering a total of $N-1$ previous consecutive feasible KPC iterations yields a total of $N$ set conditions for $f(x_0,u_0^\star)$, $N-1$ set conditions for $f(x_0,u_0^\star,u_1^\star)$, $\dots$, and $1$ set condition for $f(\dots f(f(x_0,u_0^\star),u_1^\star),\dots,u_{N-1}^\star)$. At any time $t=1,\dots,N$, the true system state is therefore contained in the intersection of the associated sets and, hence, enforcing the relaxed constraints suffices to enforce constraint satisfaction.
\bigbreak

The KPC formulation \eqref{eq.optControlProb} does not guarantee recursive feasibility due to the use of several distinct models. Nevertheless, previous successful iterations can contribute to the feasibility of future KPC problems through the stated SRS. More specifically, the closed-loop data of up to $N-1$ past steps\footnote{If only $M<N-1$ previous iterations were feasible, \eqref{eq.SRS} can be adapted to have less set intersections.} can be used to reduce the uncertainty regarding the location of true next states \textit{without updating the} KRR \textit{nominal models}. We highlight that the last constraint $x_N \in \mathbb{X}_{\text{safe}}$ is not relaxed by Proposition~\ref{prop.SRS}, but remains unchanged. 

\section{Experiments}

We illustrate the use of KPC, implemented in a receding-horizon fashion, in two different scenarios. The optimization problems were formulated with the aid of CasADi \citep{CASADI}, the Multi-Parametric Toolbox \citep{MPT}, and solved with IPOPT \footnote{Additional details about the simulations are available at \texttt{https://github.com/emilioMaddalena/KPC.}}.

\textbf{Example 1:} Consider a continuous stirred-tank reactor (CSTR) whose continuous-time dynamics are given by the differential equations
\begin{subequations}
\begin{align}
    \dot c_A(t) &= u(t) (c_{A0}-c_A(t)) - \rho_1 c_A(t) - \rho_3 \, c_A(t)^2 \\ 
    \dot c_B(t) &= -u(t) \, c_B(t) + \rho_1 c_A(t) - \rho_2 \, c_B(t)^2 
\end{align}
\end{subequations}
where $c_A$ and $c_B$ denote respectively the concentrations of cyclopentadiene and cyclopentenol, and $u$ represents the feed inflow of cyclopentadiene. We assume that the reactor temperature is constant and simulate the dynamics with the parameter values: $\rho_1 = \rho_2 = 4.1 \times 10^{-3}\text{ h}^{-1}, \rho_3 = 6.3 \times 10^{-4}\text{ h}^{-1}, c_{A0} = 5.1 \text{ mol/l}$. The constraint sets are $\mathbb{X} = \{x \in \mathbb{R}^2 \, | \begin{pmatrix} 1 & 0.5 \end{pmatrix}^\top \leq x \leq \begin{pmatrix} 3 & 2 \end{pmatrix}^\top \}$, $\mathbb{U} = \{u \in \mathbb{R} \, | \, 3 \leq u \leq 35 \}$, and the safe set is the singleton $\mathbb{X}_{\text{safe}} = \{x_{\text{eq}}\}, x_{\text{eq}} = \begin{pmatrix} 2.14 & 1.09 \end{pmatrix}^\top$ with $u_{\text{eq}} = 14.19$. The sampling period is $30$ seconds and the prediction horizon was chosen to be $N=3$. Three distinct random datasets were collected with $300$, $400$ and $500$ points respectively for the one-step, two-step and three-step ahead predictors. The noise affecting our samples was drawn randomly with uniform bound $ 1 \times 10 ^{-3}$. Squared-exponential kernels were chosen and their hyperparameters were adjusted until good fits were obtained; specifically, the lengthscales were set to larger values when dealing with higher-dimensional feature spaces. The exact regressor norms were calculated and an augmentation factor of $150\%$ was used to obtain estimates $\Gamma$. This latter step accounts for the ground-truth complexity in unexplored regions of the space. Finally, standard quadratic stage and terminal costs were employed with positive definite weight matrices. 

\begin{figure}[t]
    \floatconts
    {fig:example}% label
    {\vspace{-20pt} \caption{Left: Closed-loop system trajectories starting from different initial conditions. Right: KPC open-loop nominal predictions and uncertainty sets during all time-instants and starting from different initial conditions.}}% 
    {\includegraphics[scale=0.27]{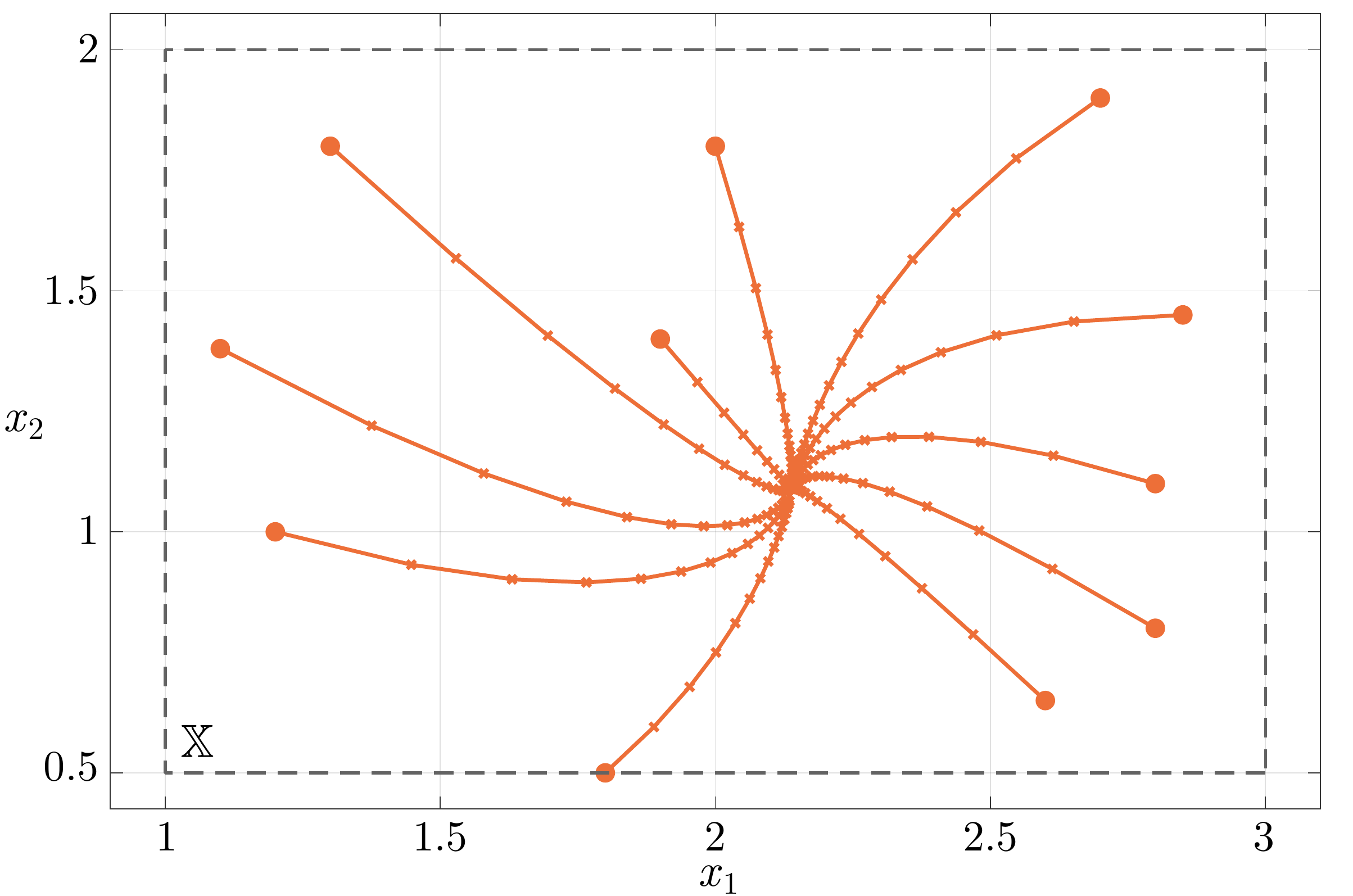}
    \hspace{1pt} \includegraphics[scale=0.27]{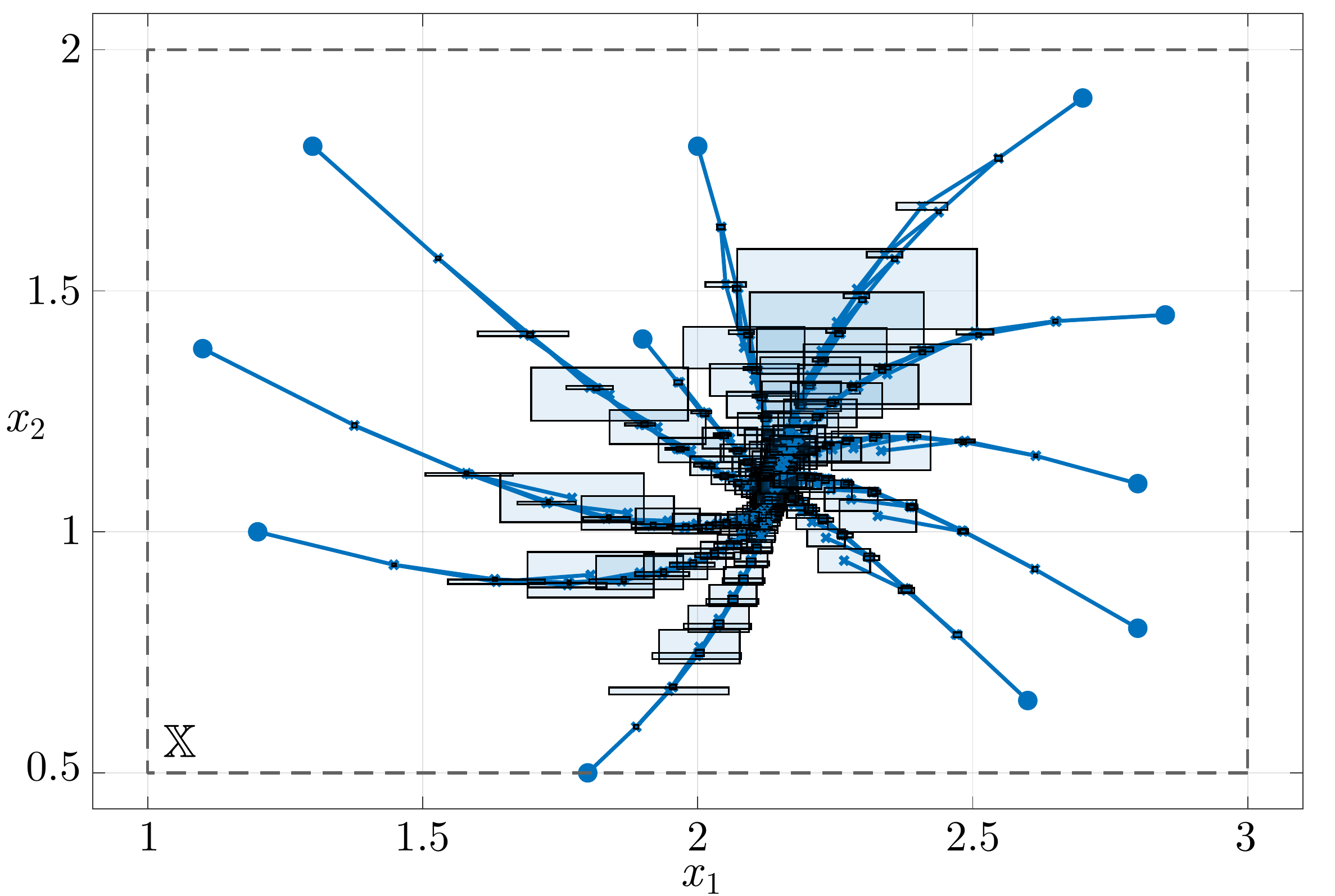}}
    \label{fig.CSTR}
\end{figure}

As is customary in practical non-linear optimal control, the terminal constraint was dropped and only a terminal penalty was employed. The system evolution starting from various initial conditions is shown in Figure~\ref{fig.CSTR}. The closed-loop trajectories (shown on the left) converged to a neighborhood of $x_S$, while all predictions and confidence sets (shown on the right) remained inside the feasible set $\mathbb{X}$ at all time-instants. It is also possible to note how predicting further into the future is more challenging as the lengths of the boxes tended to be larger at the end of the prediction horizon. 
% %The reaction speeds depend on the reactor temperature through Arrhenius' law... in the KPC formulation $Q = \text{diag}\left( 0.2, \, 1\right)$, $R = 0.5$, and $P = \begin{pmatrix} 14.46 & 13.56 \\ 13.56 & 62.22 \end{pmatrix}$.

\begin{figure}[t]
    \floatconts
    {fig:example}% label
    {\vspace{-20pt} \caption{Left: Nominal kernel-MPC predictions (\textcolor{bordeaux}{\textbf{- -}}) and closed-loop trajectory (\textcolor{bordeaux}{\textbf{|}}). Center: KPC predictions (\textcolor{ripeOrange}{\textbf{- -}}) and closed-loop trajectory (\textcolor{ripeOrange}{\textbf{|}}). Right: KPC full open-loop predictions, confidence intervals (\textcolor{azzurro}{\textbf{|}}) as error bars, and initial points depicted as circular markers. The nominal kernel-MPC predictions satisfy the constraints, but the real system violates them. KPC leads to a safe operation without constraint violations.}}% 
    {\includegraphics[scale=0.28]{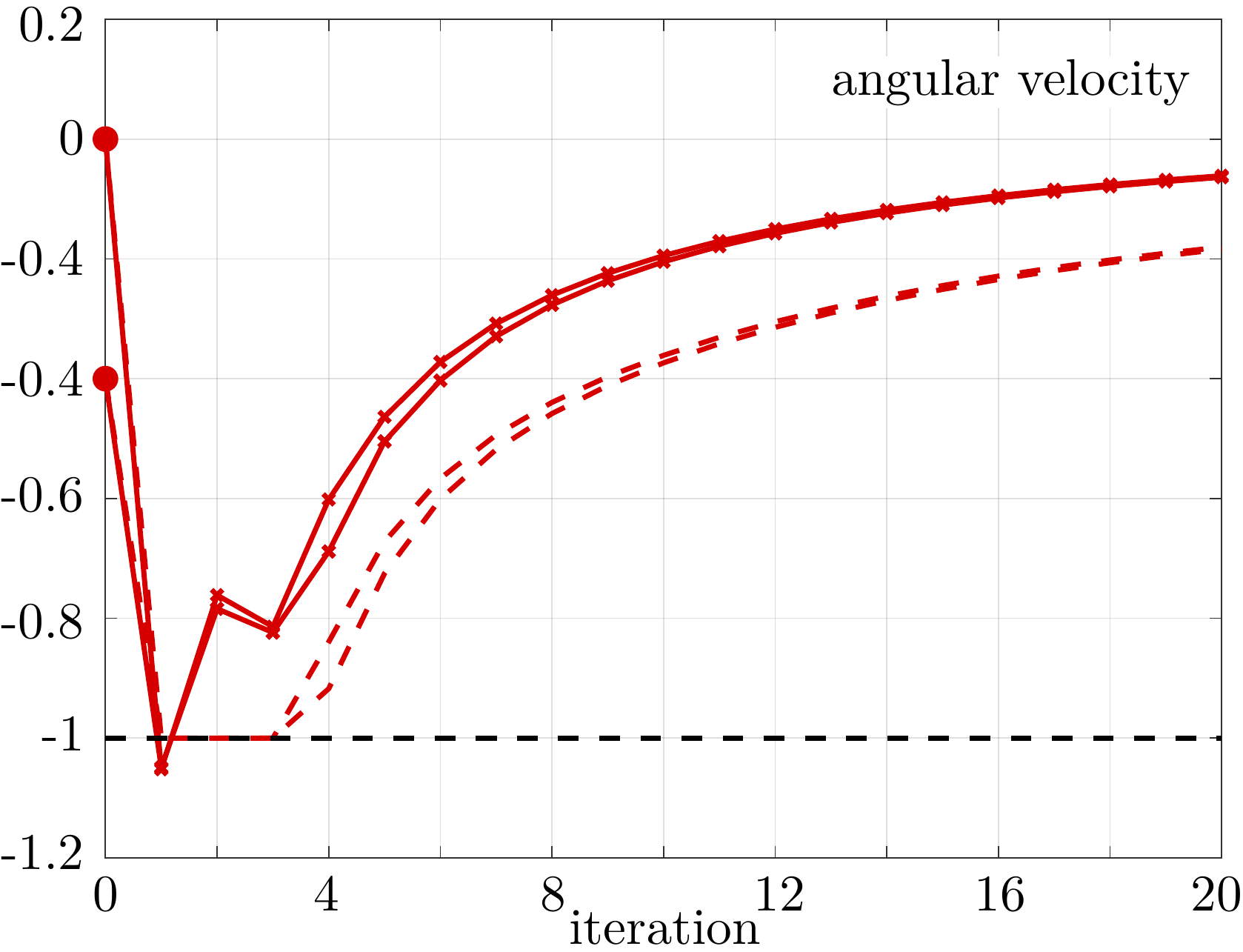}
    \hspace{1pt} \includegraphics[scale=0.28]{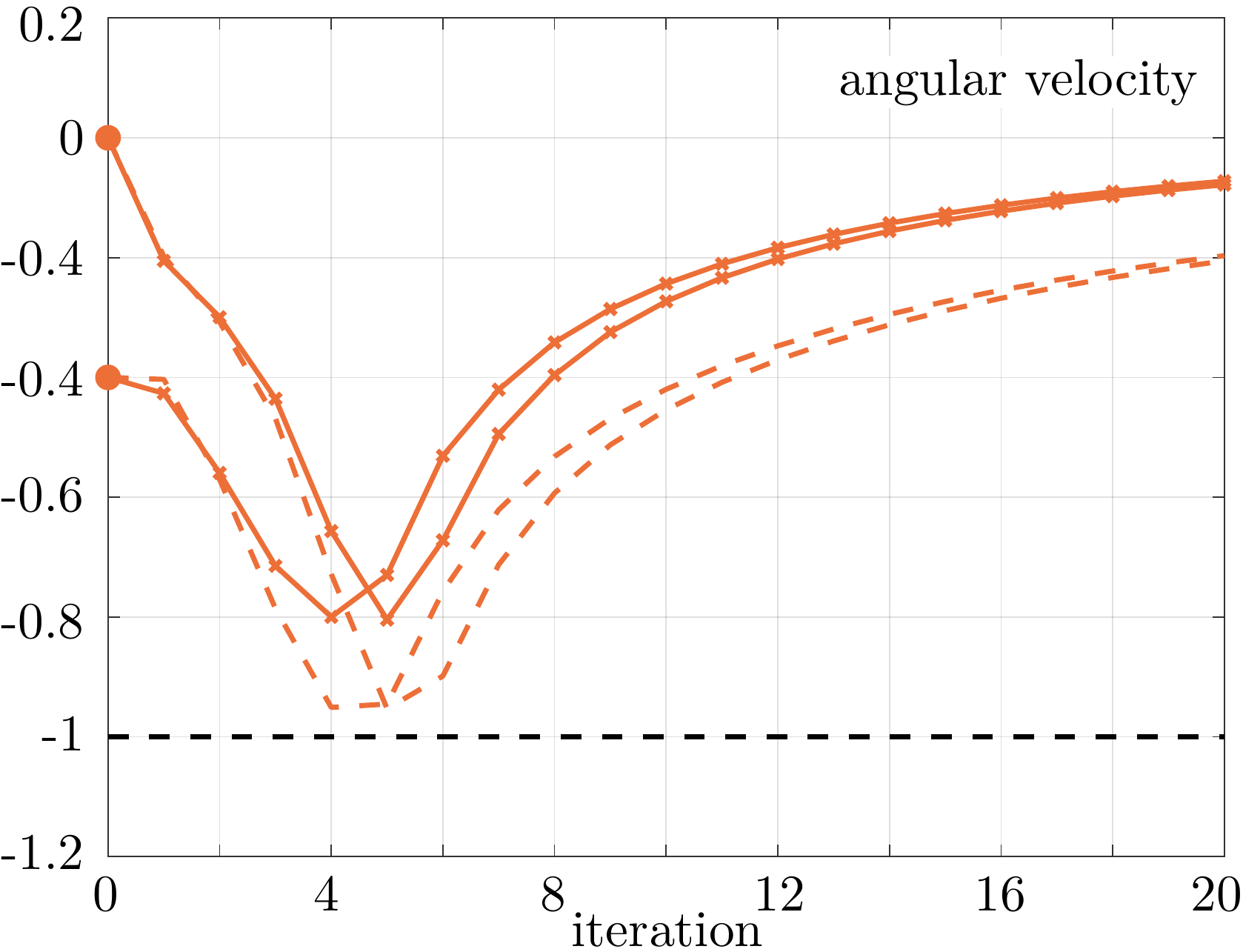}
    \hspace{1pt} \includegraphics[scale=0.28]{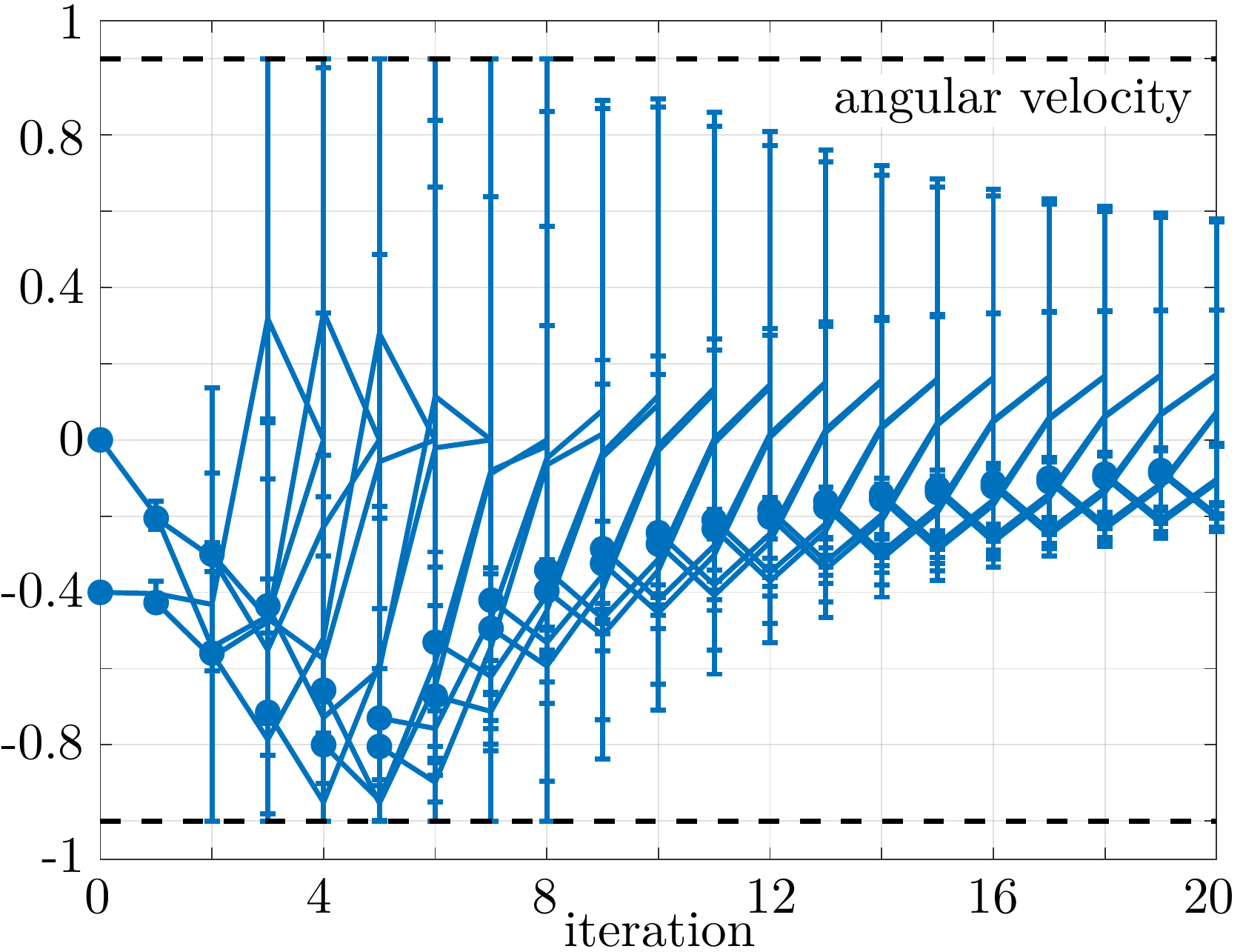}}
    \label{fig.PENDULUM}
\end{figure}

\textbf{Example 2:} The continuous-time angular dynamics of a pendulum with a rigid rod can be described by $\dot x_1 = x_2$, $\dot x_2 = (g/l)\sin(x_1) - (\nu/(ml^2)) x_2 + (1/(m l^2)) u$, where $x_1$ is its angular position, $x_2$ its angular velocity, and $u$ the torque applied to it. The parameters are: $m = 0.15$ the mass of the pendulum, $l = 0.5$ the length of the rod, $g = 9.81$ the gravitational constant, and $\nu = 0.1$ a constant for the friction model. Let the constraints be $\vert x_1 \vert \leq 3$, $\vert x_2 \vert \leq 1$, and the input be limited to $\vert u \vert \leq 1$. We selected a prediction horizon of $N=4$ and collected $D=100$ uniformly random data-points for each of the eight regression tasks. The noise was drawn randomly from a uniform distribution with bound $\bar \delta = 0.01 \, \textbf{1}$, where $\textbf{1} \in \mathbb{R}^D$ is a vector of ones. Similarly to the previous example, we used a squared-exponential kernel with increasing lengthscales and employed an augmentation factor of $300\%$ on the nominal predictor norms to estimate the $\Gamma$ constants. The sampling and control period was $0.2$~seconds. We compared KPC against \textit{nominal} kernel-MPC, i.e., a certainty equivalence approach where the state-constraints were imposed directly on the nominal predictions \eqref{eq.KPCnominal}. In the latter case, uncertainty was not quantified and the confidence sets were not present. The cost used in both formulations was a positive definite function of the states and control inputs, and included a terminal penalty term. 

Predictions and the system angular velocity evolution from two different initial conditions are shown in Figure~\ref{fig.PENDULUM}. As can be seen from the plots, imposing the state-constraints on the nominal model predictions was not sufficient to guarantee safety as the closed-loop system behavior violated the $x_2 \geq -1$ restriction. On the other hand, since KPC quantified and incorporated the associated uncertainty into the optimization problems, the constraints were satisfied. The error bars on the right plot show the all predictions and uncertainty values at each step in the form of error bars. Note that the safety constraints were active at multiple points in time. 

\section{Concluding Remarks}

KPC was proposed as a predictive control methodology based on non-parametric kernel models and their associated uncertainty estimates. Its key feature is deterministic constraint satisfaction when a solution to the optimization problem is found. From an approximation theory perspective, future works could study the advantages of employing SVR surrogate models over KRR ones, as well as refining the existing error-bounds, which we believe to be possible. Establishing conditions under which KPC would enjoy additional closed-loop properties such as convergence is deemed as interesting and could guide practical real-world applications of the proposed control scheme. 

%%\begin{itemize}
%%  \item Limit the main text (not counting references) to 10 PMLR-formatted pages, using this template.
%%  \item Include {\em in the main text} enough details, including proof details, to convince the reviewers of the contribution, novelty and significance of the submissions.
%%\end{itemize}

% Acknowledgments---Will not appear in anonymized version
\acks{This work received support from the Swiss National Science Foundation under the Risk Aware Data-Driven Demand Response project (grant number 200021 175627) and CSEM's Data Program.}

\bibliography{refs}

\end{document}